\documentclass[11pt,preprintnumbers,aps,amssymb,nofootinbib,amsmath]
{revtex4}
\usepackage{epsfig,epsf}
\usepackage{graphicx}
\usepackage{verbatim}
\usepackage{epstopdf}
\usepackage{bm} 
\newcommand{\beq}{\begin{equation}}
\newcommand{\beql}[1]{\begin{equation}\label{#1}}
\newcommand{\eeq}{\end{equation}}
\newcommand{\bea}{\begin{eqnarray}}
\newcommand{\eea}{\end{eqnarray}}

\def\eq#1{{(\ref{#1})}}
\def\fig#1{{Fig.~\ref{#1}}}

\def\sec#1{{Sec.~\ref{#1}}}
\newcounter{topiccounter}
\newcommand{\as}{\alpha_s}
\newcommand{\bas}{\bar\alpha_s}

\def\b#1{\mathbf{#1}}

\begin{document}

\preprint{RBRC-763}

\title{Coherent and incoherent diffractive  hadron production in $pA$ collisions and gluon saturation}

\author{Kirill Tuchin$\,^{a,b}$\\}

\affiliation{
$^a\,$Department of Physics and Astronomy, Iowa State University, Ames, IA 50011\\
$^b\,$RIKEN BNL Research Center, Upton, NY 11973-5000\\}

\date{\today}


\begin{abstract}
We study coherent and incoherent diffractive hadron production in high energy quarkonium--heavy nucleus  collisions as a  probe of the gluon saturation regime of QCD. Taking this process as a model for $pA$ collisions, 
we argue that the coherent diffractive gluon production, in which the target nucleus stays intact, exhibits a remarkable sensitivity to the energy, rapidity and atomic number dependence. The incoherent diffractive gluon production is less sensitive to the details of the low-$x$ dynamics but can serve as a probe of fluctuations in the color glass condensate.  As a quantitative measure of the 
nuclear effects on diffractive hadron production we introduce a new observable -- the diffractive nuclear modification factor. 
We discuss  possible  signatures of gluon saturation in diffractive gluon production at RHIC, LHC and EIC.

\end{abstract}

\maketitle

\section{Introduction}\label{sec:intr}

Coherent diffractive hadron production in $pA$ collisions is a process $p+A\to X+h+[LRG]+A$, where $[LRG]$ stands for Large Rapidity Gap. We have
recently argued  in \cite{Li:2008bm,Li:2008jz,Li:2008se} that the coherent diffractive hadron production can serve as a sensitive probe of the low-$x$ dynamics of the nuclear matter. The coherent diffractive production exhibits a much stronger dependence on energy and atomic number  than the corresponding inclusive process. Indeed, the diffractive amplitude is proportional to the square of the inelastic one. At \emph{asymptotically} high energies the coherent diffractive events are expected to constitute a half of the total cross section, other half being all inelastic processes. Therefore, coherent diffraction  is a powerful tool 
for studying the low-$x$ dynamics of QCD. In particular, we advocated using  the coherent diffraction as a tool for studying the gluon saturation \cite{Gribov:1983tu,Mueller:1986wy}. 

The low-$x$ region of QCD is characterized by strong gluon fields \cite{McLerran:1993ni,Kovchegov:1996ty} and can be described in the framework of the color glass condensate\cite{Balitsky:1995ub,JalilianMarian:1997jx,Jalilian-Marian:1997dw,
Kovner:2000pt,Iancu:2000hn,Ferreiro:2001qy}. Equations describing the color glass condensate take the simplest form in the  the mean-field approximation. It is valid for a color field of a heavy nucleus in the multicolor limit; the relevant resumation parameter is $\as^2A^{1/3}\sim 1$. To the leading order in this parameter, the nuclear color field  is a coherent non-Abelian  superposition of the color fields of single nucleons. Higher order corrections arise due to the nucleon--nucleon or parton-parton correlations within a nucleon. The later contribute towards the gluon saturation in proton.  

In all phenomenological applications of the CGC formalism, one usually relies on the   mean-field approximations in which only the lowest order  Green's functions are relevant. If rapidity interval $Y=\ln (1/x)$ is such that $\as Y<1$, then the low-$x$ quantum evolution effects are suppressed and the color field can be treated quasi-classically. When $\as Y\sim 1$, the quantum corrections become important and are taken into account by Balitsky-Kovchegov equation \cite{Balitsky:1995ub,Kovchegov:1999yj}, which is the first (truncated) equation in the infinite hierarchy of  the coupled integro-differential equations governing evolution of Green's functions of all orders (JIMWLK equations \cite{Balitsky:1995ub,JalilianMarian:1997jx,Jalilian-Marian:1997dw,
Kovner:2000pt,Iancu:2000hn,Ferreiro:2001qy}).

Although corrections to the mean-field approximation, i.e.\ quantum fluctuations about the classical solution,  are assumed to be small in $pA$ collisions at RHIC their detailed phenomenological study is absent. In this paper we fill this gap by calculating the differential  cross section for the incoherent diffractive gluon production which happens to be proportional to the dispersion of the quasi-classical scattering amplitude in the impact parameter space, as is expected in the Glauber theory \cite{Glauber:1955qq}. The incoherent diffractive gluon production in $pA$ collisions is a process 
\beq\label{id}
p+A\to X+h+[LRG]+A^*\,, 
\eeq
 where $A^*$ denotes the excited nucleus that subsequently decays into a system of colorless protons, neutrons and nuclei debris.  

Note, that the incoherent diffraction \eq{id} measures fluctuations of the nuclear color field. At higher energies color field of proton also saturates. The mean-field approximation is not at all applicable in this case. A quantum fluctuations (e.g.\ the so-called ``pomeron loops" \cite{Iancu:2004iy,Iancu:2004es,Kovner:2005en,Kovner:2007zu}) may be the driving force of the gluon saturation   in proton. This problem is of great theoretical and phenomenological interest. However, in spite of considerable theoretical efforts solution to this problem is still illusive (see \cite{Kovner:2007zz} for a brief review of recent advances). 

Motivated by theoretical and phenomenological interest to the incoherent diffraction as a measure of fluctuations of the saturated nuclear color field, we derive a formula for the incoherent diffractive hadron production \eq{id} in the framework of the color glass condensate using the dipole model of Mueller \cite{dip} and use it for numerical study in the RHIC and LHC kinematic regions. 
The limiting cases of large and small invariant masses of the diffractively produced system were previously considered by many authors \cite{Wusthoff:1997fz,Bartels:1999tn,Kopeliovich:1999am,Kovchegov:2001ni,GolecBiernat:2005fe,Marquet:2004xa,Marquet:2007nf,Munier:2003zb}. Inclusion of the diffractive gluon production was recently shown to be essential for the phenomenology of diffractive DIS off heavy nuclei \cite{Kowalski:2008sa}.
We generalize there results for all invariant masses (satisfying  \eq{cL2}) and include the gluon evolution effects at \emph{all} rapidity intervals.  Unlike models based on the leading twist nuclear shadowing \cite{Frankfurt:2003zd,Guzey:2005ys}, we explicitly take into account the gluon saturation effects in the nucleus.

The paper is structured as following. In \sec{sec:glauber} we review the Glauber approach \cite{Glauber:1955qq} to the coherent and incoherent diffraction in $pA$ collisions. In \sec{sec:dipole} we discuss generalization of this approached to the case of quarkonium--nucleus scattering in high energy QCD. We then turn in \sec{sec:dgp-qc} to the main subject of the paper -- the diffractive hadron production -- and calculate contributions from coherent and incoherent channels in the quasi-classical approximation neglecting the low-$x$ evolution effects. The low-$x$ evolution effects are taken into account in \sec{sec:lowx}. 

Numerical results of the diffractive cross section are presented in \sec{sec:nmf}.  
Our phenomenological analyses indicates that the kinematic regions of RHIC and LHC are \emph{not} asymptotic as far as the energy dependence of the diffractive hadron production  is concerned. We observed that the ratio of the coherent and incoherent diffractive hadron production at midrapidity in $pA$ collisions  increases from RHIC to LHC. This is because  gluon  saturation effects in proton are assumed to be small. Absence of this feature in experimental measurements  can serve as an evidence for the onset of gluon saturation in proton.  

It is important to emphasize that all our results are obtained without imposing any experimental constraints on the forward scattering angle measurements. 
Coherent diffraction measurements require much better forward scattering angle resolution than the incoherent ones. Therefore,  incoherent diffraction may well turn out to be the only one accessible in experiment.  

We summarize in \sec{sec:summary}.

\section{Diffraction in the Glauber Model}\label{sec:glauber}

\subsection{Total cross section for coherent diffraction in $pA$}

 A general approach to multiple scattering in high energy nuclear physics was suggested by Glauber \cite{Glauber:1955qq}. First, consider the $pp$ scattering and introduce the elastic $pp$ scattering amplitude $i\Gamma^{pp}(s,\b b)$, where $s$ is the center-of-mass energy squared and $\b b$ is an impact parameter. At high energies $\b b$ is a two-dimensional vector in the transverse plane. According to the optical theorem,
\beq\label{optic.pp}
\sigma_\mathrm{tot}^{pp}=2\int d^2b \, \mathrm{Re}\,\Gamma^{pp}(s,\b b )\,.
\eeq
The scattering amplitude $\Gamma^{pp}(s,\b b)$ can be written as
\beq\label{phase1}
\Gamma^{pp}(s,\b b)=1-e^{-i\chi^{pp}(s,\b b )}\,,
\eeq
where $\chi^{pp}(s,\b b )$ is a change of phase due to interaction at point $\b b$.
In the following we will always assume the isotopic invariance of the scattering amplitudes, e.g.\ $\Gamma^{pp}(s,\b b)= \Gamma^{pN}(s,\b b )$, where $N$ stands for a nucleon.

In $pA$ collisions  momenta of nucleons can be neglected as compared to the incoming proton momentum. Therefore, their positions $\b b_a$, $a=1,\ldots, A$ are fixed during the interaction. If scattering of proton on a different nucleons is independent, then the corresponding phase shifts $\chi_a^{pN}(s,\b B-\b b_a)$ add up:
\beq\label{phase.pa}
\chi^{pA}(s,\b B, \{ \b b_a\})= \sum_{a=1}^A \chi_a^{pN}(s,\b B-\b b_a)\,,
\eeq
where $\b B$ is proton's impact parameter.  Indeed, this result holds in QCD as was demonstrated by A.~Mueller \cite{Mue}, see \sec{sec:dipole}. The scattering amplitude of proton $p$ on a nucleus $A$ reads
\beq\label{gamma.pa}
\Gamma^{pA}(s,\b B, \{ \b b_a\})= 1-e^{-i\chi^{pA}(s,\b B, \{ \b b_a\})}=
1-\prod_{a=1}^Ae^{-i\chi^{pN}_a(s,\b B-\b b_a )}=1-\prod _{a=1}^A\left(
1-\Gamma^{pN}(s,\b B-\b b_a)\right)\,.
\eeq

Introduce the  nucleus-averaged amplitude 
\beq\label{gii1}
\Gamma^{pA}_{if}(s,\b B )=\langle A_i|\Gamma^{pA}(s,\b B, \{\b b_a\} )|A_f\rangle\,,
\eeq
with $|A_i\rangle$ being the initial  and $|A_f\rangle$ the final nucleus state. Then, by the optical theorem,  the total $pA$ cross section is given by  
\beq\label{optic.pa}
\sigma_\mathrm{tot}^{pA}=2\int d^2b \,\mathrm{Re}\, \Gamma^{pA}_{ii}(s,\b b )\,.
\eeq

Since we neglect motion of nucleons during the interaction, the
distribution of nucleons in the nucleus is completely specified by the thickness function $T(\b b)$:
\beq\label{thickness}
 T(\b b)=\int_{-\infty}^\infty\, dz\, \rho_A(\b b,z)\,,
\eeq
where $\rho_A(\b b,z)$ is nuclear density and $z$ is the longitudinal coordinate.  The thickness function is often written in terms of the mean density of nucleons $\rho= A/(\frac{4}{3}\pi R_A^3)$ and the  nuclear profile function $T_A(\b b)$ as 
\beq\label{t2t}
T(\b b)=\rho\,T_A(\b b)\,.
\eeq
It is normalized as 
\beq\label{nucl.den}
\int d^2b\,T(\b b)=\int d^2b\,\rho\, T_A(\b b)=A\,.
\eeq

Using these definitions  we write the diagonal element of \eq{gii1} as 
\bea\label{glaub1}
\Gamma^{pA}_{ii}(s,\b B )&=&\frac{1}{A}\int \prod_{a=1}^A\,d^2b_a\, \rho\, T_A(\b b_a)\, \Gamma^{pA}(s,\b B-\b b_a )\nonumber\\
&=&1-\left[ 1-\frac{1}{A}\int d^2b_a\, \Gamma^{pN}(s, \b b_a)\, \rho\, T_A(\b B-\b b_a)\right]^A\nonumber\\
&\approx& 1-e^{-\int d^2b_a\, \Gamma^{pN}(s, \b b_a)\, \rho\, T_A(\b B-\b b_a)}\,,
\eea
where we assumed that $A\gg 1$. 

Impact parameter dependence of the $pp$ collisions is usually parameterized as
\beq\label{bpp}
\Gamma^{pp}(s,\b b)=\frac{1}{2}\sigma_\mathrm{tot}^{pp}(s)\,S_p(\b b)\,,
\eeq
where we neglected a small imaginary part of $\Gamma^{pp}(s,\b b)$ and 
introduced the proton profile function $S_p(\b b)$ as
\beq\label{sp}
S_p(\b b)= \frac{1}{\pi R_p^2}\,e^{-\frac{b^2}{R_p^2}}\,.
\eeq
Phenomenologically, $R_p=\sqrt{2B}\approx 1$~fm, where $B=12.6$~GeV$^{-1}$. Since the proton radius  $R_p$ is much smaller than the radius of a heavy nucleus $R_A$,  we can approximate the proton profile function by the delta-function in which case the overlap integral appearing in \eq{glaub1} becomes simply the nuclear thickness function
\beq\label{overlap}
\int d^2b\, S_p(\b b)\,\rho\, T_A(\b B-\b b)\approx \rho\, T_A(\b B)\,.
\eeq
This approximation holds when $R_p\ll R_A$, i.e.\ $A^{1/3}\gg 1$. Actually, this condition follows from a requirement that  the quasi-classical approximation holds. Indeed, in the quasi-classical approximation 
 $\as\ll 1$ and $\as^2 A^{1/3}\sim 1$ implying that $A^{1/3}\gg 1$.
 
Using \eq{bpp} and \eq{overlap} in \eq{glaub1} we derive
\beq\label{glaub2}
\Gamma^{pA}_{ii}(s,\b b )= 1-e^{-\frac{1}{2}\sigma^{pN}_\mathrm{tot}(s)\, \rho\, T_A(\b b)}\,.
\eeq
The cross section of coherent diffraction is given by  the elastic cross section
\beq\label{cd2}
\sigma_\mathrm{CD}^{pA}(s)= \int d^2b \, \left|\Gamma^{pA}_{ii}(s,\b b )\right|^2=
\int d^2b \left( 1-e^{-\frac{1}{2}\sigma^{pN}_\mathrm{tot}(s)\, \rho\, T_A(\b b)}\right)^2 \,.
\eeq

\subsection{Total cross section for incoherent diffraction in $pA$}

Consider  incoherent diffraction of proton $p$ on a nucleus $A$.  In this processes the nucleus gets excited from the initial state  $|A_i\rangle$ to any colorless final state $|A_f\rangle$. It may then decay, but it is essential that its constituent nucleons remain intact as color objects. 
The corresponding cross section reads
\bea\label{aa1}
\sigma^{pA}_\mathrm{ID}(s)&=& \int d^2B\sum_{i\neq f}\langle A_f|\Gamma^{pA}(s,\b B, \{\b b_a\})|A_i\rangle ^\dagger
\langle A_f|\Gamma^{pA}(s,\b B, \{\b b_a\})|A_i\rangle\nonumber\\
&=& \int d^2B\sum_{f}\langle A_f|\Gamma^{pA}(s,\b B, \{\b b_a\})|A_i\rangle ^\dagger
\langle A_f|\Gamma^{pA}(s,\b B, \{\b b_a\})|A_i\rangle \nonumber\\ 
&&\,-\,
\int d^2B  \left|\langle A_i|\Gamma^{pA}(s,\b B, \{\b b_a\})|A_i\rangle\right|^2\nonumber\\
&=& \int d^2B \left[ \langle A_i|\left|\Gamma^{pA}(s,\b B, \{\b b_a\})\right|^2|A_i\rangle-\left|\langle A_i|\Gamma^{pA}(s,\b B, \{\b b_a\})|A_i\rangle\right|^2\right]\,
\eea
where in the last line we used the completeness of the set of nuclear states. 
We arrived at the well-known result that the cross section for the incoherent diffraction at a given impact parameter $\b B$ is given by the square of the standard deviation of  the scattering amplitude $\Gamma^{pA}(s,\b B, \{\b b_a\})$ from its mean-field value $\langle A_i|\Gamma^{pA}(s,\b B, \{\b b_a\})|A_i\rangle$ in a space span by  impact parameters $\{\b b_a\}$. Clearly, in the black-disk limit, corresponding to the asymptotically high energies $s\to \infty$,  this deviation vanishes because $\Gamma^{pN}(s,\b B, \{\b b_a\})\to 1$ for $|\b B- \b b_a|<R_A$ and is zero otherwise, see \eq{gamma.pa}. The standard deviation is a measure of quantum fluctuations near the mean-field value.  

Since, $\Gamma^{pN}(s,\{\b b_a\})$ is approximately real, we have using \eq{gamma.pa}
\bea\label{aa2}
\left( \Gamma^{pA}(s,\b B, \{\b b_a\}\right)^2&= &\left( 1-\prod_{a=1}^A\left[ 1- \Gamma^{pN}(s,\b B-\b b_a)\right]\right)^2\nonumber\\
&=& 1-2\prod_{a=1}^A\left[  1- \Gamma^{pN}(s,\b B-\b b_a)\right]+
\prod_{a=1}^A \left[  1- \Gamma^{pN}(s,\b B-\b b_a)\right]^2
\eea
Averaging over the nucleus and taking the large $A$ limit we obtain 
\bea\label{aa3}
&&
\langle A_i|\left|\Gamma^{pA}(s,\b B, \{\b b_a\})\right|^2|A_i\rangle=\nonumber\\
&&
1-2\,e^{-\int d^2b\, \Gamma^{pN}(s,\b B-\b b)\rho\, T_A(b)}+
e^{-\int d^2b\, \left[2\Gamma^{pN}(s,\b B-\b b)-(\Gamma^{pN}(s,\b B-\b b))^2\right] \rho\, T_A(b)}\nonumber\\
&&= 1-2\, e^{-\frac{1}{2}\sigma_\mathrm{tot}^{pN}(s)\,\rho\, T_A(\b b)}+e^{-\sigma_
\mathrm{in}^{pN}(s)\,\rho\, T_A(\b b)} \,,
\eea
where we used  \eq{bpp} and denoted the inelastic  $pN$ cross section as $\sigma_\mathrm{in}^{pN}(s)$.  Substituting \eq{aa3} into \eq{aa1} and using \eq{cd2} we derive
\beq\label{id2}
\sigma_\mathrm{ID}^{pA}(s)=\int d^2b\, e^{-\sigma_\mathrm{in}^{pN}(s)\,\rho\, T_A(\b b)}\,
\left[ 1- e^{-\sigma_\mathrm{el}^{pN}(s)\,\rho\, T_A(\b b)}\right]\,.
\eeq

Elastic cross section $\sigma_\mathrm{el}^{pN}$, which appears in \eq{id2}, can be found by taking square of  \eq{bpp} and integrating over the impact parameter:
\beq\label{el}
\sigma^{pN}_\mathrm{el}=\int d^2b \left|\Gamma^{pN}(s,\b b)\right|^2=
\frac{1}{4}(\sigma_\mathrm{tot}^{pN})^2\int d^2b\, S_p^2(\b b)= 
\frac{(\sigma_\mathrm{tot}^{pN})^2}{8\pi R_p^2}\,,
\eeq
where we used \eq{sp}.

It is seen from \eq{cd2} and \eq{id2} that since the $pN$ cross sections increase with energy, at asymptotically high energies the incoherent diffraction cross section vanishes, whereas the coherent one reaches a half of the total cross section. This well-known conclusion is a consequence of unitarity of the scattering amplitude and thus is independent of interaction details.

\section{Diffraction in the dipole model}\label{sec:dipole}

It is phenomenologically  reasonable to approximate the proton light-cone wave-function (away from fragmentation regions) by a system of color dipoles \cite{Avsar:2005iz,Kopeliovich:2005us,Li:2008se}. If separation of quark and anti-quark is small, one can apply the perturbation theory to calculate the scattering amplitude of quarkonium on the nucleus. It was demonstrated by Mueller in \cite{dip} that at high energies the $q\bar q A$ forward elastic scattering amplitude takes exactly the same form as \eq{glaub2} with the $pA$ cross section replaced by the $q \bar qA$ one. By virtue of the optical theorem, the total quarkonium--nucleus cross section reads 
\beq\label{sig.tot.dm}
\sigma_\mathrm{tot}^{q\bar qA}(s;\b r)=2\int d^2b \, N_A(\b r, \b b, Y)=
 2\int d^2b \, \left( 1-e^{-\frac{1}{2}\sigma^{q\bar qN}_\mathrm{tot}(s;\b r)\, \rho\, T_A(\b b)}\right)\,,
\eeq
where 
\beq\label{NA}
N_A(\b r, \b b, Y)=\mathrm{Re}\, \Gamma_{ii}^{q \bar qA}(s,\b b;\b r)
\eeq
is the imaginary part of the forward elastic $q\bar qA$ scattering amplitude and $Y=\ln(1/x)=\ln (s/s_0)$ is rapidity with $s_0$ a reference energy scale.  Eq.~\eq{sig.tot.dm} is called the Glauber--Mueller formula. Let us note that it is far from obvious that the high energy amplitude in \eq{sig.tot.dm} must have the same form as the low energy one \eq{glaub2}. In this correspondence it is crucial that the color dipoles are identified as the relevant degrees if freedom at high energies. 
Eq.~\eq{sig.tot.dm} holds when $ \ln(m_NR_A)\ll Y\ll 1/\as$, where $m_N$ is a nucleon mass. This condition guarantees that  the coherence length $l_c$ is much larger than the nuclear radius, see \eq{cL1}, and  that the low-$x$ gluon evolution is suppressed. In this case
 $q\bar q N$ cross section can be calculated in the Born approximation  (two-gluon exchange) as
\beq\label{born}
\sigma_\mathrm{tot}^{q\bar qN}(s;\b r)= \frac{\as}{N_c}\pi^2\b r^2\, xG(x, 1/\b r^2)\,,
\eeq
with the gluon distribution function 
\beq\label{xG}
xG(x,1/\b r^2)=\frac{\as C_F}{\pi}\ln \frac{1}{\b r^2\mu^2}\,,
\eeq
where $\mu$ is an infrared cutoff.  We can re-write \eq{born} in terms of the \emph{gluon} saturation scale $Q_{s0}^2$  defined as
\beq\label{sat.sc}
Q_{s0}^2= \frac{4\pi^2\as N_c}{N_c^2-1}\,\rho\, T_A(\b b)\,xG(x,1/\b r^2)\,.
\eeq
Subscript $0$ indicates that the low-$x$ evolution is suppressed. Eq.~\eq{xG} implies that in the Born approximation the $q\bar qN$ total cross section \eq{born} and hence the  saturation scale \eq{sat.sc} are energy  independent. Using this definition we have
\beq\label{rss}
\rho\, T_A(\b b)\,\sigma_\mathrm{tot}^{q\bar q N}(s;\b r)=\frac{C_F}{2N_c}\, \b r^2\, Q^2_{s0}\,,
\eeq
where $C_F=(N_c^2-1)/(2N_c)$.  Substituting \eq{rss} into  \eq{sig.tot.dm} we derive another representation of the total quarkonium--nucleus cross section
\beq\label{dm1}
\sigma_\mathrm{tot}^{q\bar qA}(s;\b r)=
 2\int d^2b \, \left( 1-e^{-\frac{1}{4} \frac{C_F}{N_c}\,\b r^2\, Q_{s0}^2}\right)\,.
\eeq

Integrand of \eq{dm1} represents the propagator of quarkonium $q\bar q$ of size $\b  b$ through the nucleus at impact parameter $\b b$.  Since $\sigma_\mathrm{tot}^{q\bar qN}\sim \as^2$ and $\rho\, T_A(\b b)\sim A^{1/3}$  Eq.~\eq{dm1} sums up powers of $\as^2 A^{1/3}$. It has been shown by Kovchegov that this corresponds to the coherent scattering of $q\bar q$ off the quasi-classical  field of the nucleus \cite{Kovchegov:1996ty}. This is possible only if the coherence length $l_c$ of the $q\bar q$ pair is much larger than the nuclear radius. This condition certainly holds at RHIC energies. In the quasi-classical approximation the coherent diffraction cross section is given by  (cp.\ \eq{cd2}) 
\beq\label{cd2.dm}
\sigma_\mathrm{CD}^{q\bar qA}(s; \b r)= \int d^2b \, \left|\Gamma^{q\bar qA}_{ii}(s,\b b; \b r )\right|^2=
\int d^2b \left( 1-e^{-\frac{1}{4}\frac{C_F}{N_c} Q_{s0}^2\, \b r^2}\right)^2 \,,
\eeq
Eq.~\eq{cd2.dm} represents the leading  contribution to the diffraction cross section in the quasi-classical field strength. On the contrary, the incoherent diffraction cross section, see\eq{id2}, vanishes in this mean-field approximation since $\sigma_\mathrm{el}^{q\bar qN}\rho\, T_A(\b b)\sim \as^4 A^{1/3}\sim \as^2\ll 1$.  Vanishing of the incoherent diffraction cross section can also be seen directly from \eq{aa1}. 
It is a consequence of vanishing relative fluctuations at large occupation numbers of classical fields. This effect has recently been studied within the Glauber framework (see \sec{sec:glauber}) in \cite{Kaidalov:2003vg,Guzey:2005tk,Kopeliovich:2005us}.

Let us emphasize to avoid a possible confusion, that the semantic distinction between the coherent and incoherent diffraction concerns the nucleus target staying intact or breaking down in the collision. As far as  the coherence length is concerned, at high enough energies such that $l_c\gg R_A$ the \emph{scattering} is coherent for both coherent and incoherent diffraction. However, unlike the incoherent diffraction that can happen in the incoherent scattering at low energies (provided only that $l_c\gg R_N$), the coherent diffraction is possible only if the scattering is coherent over the entire nucleus.

\section{Diffractive gluon production}\label{sec:dgp-qc}
   
 \subsection{Coherent diffractive gluon production}  
Consider coherent diffractive production of a gluon of momentum $k$ in $q\bar qA$ collision. As mentioned before, coherent diffraction is possible only if the coherence length $l_c$ of the emitted gluon with momentum $k$ is larger than the nucleus size $R_A$ (in the nucleus rest frame):
\beq\label{cL1}
l_c=\frac{k_+}{\b k^2}\gg R_A\,,
\eeq
where + indicates the light-cone direction of the incoming proton. The invariant mass of the produced system is given by $M^2=\b k^2/x$, where $x=k_+/p_+$ and $p$ is the proton momentum. Substituting these equations in \eq{cL1} yields  the following condition on the mass of the diffractive system:
\beq\label{cL2}
M^2\ll \frac{p_+}{R_A}=\frac{s}{R_Am_p}\,,
\eeq
where $\sqrt{s}$ is the center-of-mass energy of the proton--nucleon collision and $m_p$ is  proton mass.

Coherent diffractive gluon production off the large nucleus was calculated in \cite{Kovchegov:2001ni,Kovner:2001vi,Kovner:2006ge}. 
The corresponding cross section reads
\bea\label{xsectQ}
\frac{d\sigma_\mathrm{CD}(k,y)}{d^2k\, dy}&=&\,\frac{1}{(2\pi)^2}\,\int d^2b\, d^2z_1\,
d^2z_2\,\Phi^{q\bar q}(\b x, \b y, \b z_1, \b z_2)\,
e^{-i\b k\cdot(\b z_1-\b z_2)}\,\nonumber\\
 &&\times\,
 \left( \Gamma^{q\bar q GA}_{ii}(s,\b b;\b x,\b y,\b z_1)-\Gamma^{q\bar q A}_{ii}(s,\b b;\b x,\b y)\right)\nonumber\\
 &&\times\,
  \left( \Gamma^{q\bar q GA}_{ii}(s,\b b;\b x,\b y,\b z_2)-\Gamma^{q\bar q A}_{ii}(s,\b b;\b x,\b y)\right)
  \,,
\eea
where  the $q\bar q \to q\bar q G$ light-cone wave function $\Phi^{q\bar q}$ is given by 
\beq\label{wf}
\Phi^{q\bar q}(\b x, \b y, \b z_1, \b z_2)= \frac{\as
C_F}{\pi^2}\,\left(\frac{\b z_1-\b x}{|\b z_1-\b x|^2}-
 \frac{\b z_1-\b y}{|\b z_1-\b y|^2}\right)\cdot
 \left(\frac{\b z_2-\b x}{|\b z_2-\b x|^2}-
 \frac{\b z_2-\b y}{|\b z_2-\b y|^2}\right)\,.
\eeq
 The $q\bar q$ scattering amplitude $\Gamma^{q\bar q A}_{ii,\sigma}(s,\b b)$ is given by \eq{cd2.dm}, while the $q\bar q G$ one was calculated in 
\cite{Kopeliovich:1998nw,Kovner:2001vi,Tuchin:2004rb} and reads
\bea\label{qqG}
\Gamma^{q\bar q GA}_{ii,\sigma}(s,\b b;\b x, \b y, \b z_\sigma)&=& 
\langle A_i| \Gamma^{q\bar qG A}_{\sigma}(s,\b B, \{\b b_a\};\b x, \b y, \b z_\sigma) |A_i\rangle =
1-e^{-\frac{1}{2}\sigma_\mathrm{tot}^{q\bar qGN}(s;\b x, \b y, \b z_\sigma)\,\rho\, T_A(\b b)}  
\nonumber\\
&=&1-e^{-\frac{1}{8}(\b x-\b z_\sigma)^2Q_{s0}^2
-\frac{1}{8}(\b y-\b z_\sigma)^2Q_{s0}^2-\frac{1}{8N_c^2}(\b x-\b y)^2Q_{s0}^2}\,,
\eea
where $\b x$, $\b y$ and $\b z_\sigma$ are the transverse coordinates of quark, antiquark and gluon respectively; $\sigma =1$ in the amplitude and $\sigma=2$  in the complex conjugated one. For future reference, note that we can express the $q\bar q GN$ total cross section in terms of the $q\bar qN$ one given by \eq{rss}
\beq\label{expr}
\sigma_\mathrm{tot}^{q\bar qGN}(s;\b x, \b y, \b z_\sigma)= \frac{N_c}{2C_F}\left[ 
\sigma_\mathrm{tot}^{q\bar qN}(s;\b x-\b z_\sigma )+
\sigma_\mathrm{tot}^{q\bar qN}(s;\b y-\b z_\sigma )-
\frac{1}{N_c^2}\sigma_\mathrm{tot}^{q\bar qN}(s;\b x-\b y)\right]\,,
\eeq
where we explicitly indicated the dipole size and energy dependence of the $q\bar q N$ total cross section.

Depending on the relation between the gluon emission time $\tau_\sigma$ and the interaction time $\tau'_\sigma$ in the amplitude and in the c.c.\ amplitude  there are four possible products  of the $q\bar q$ and $q\bar q G$ amplitudes appearing in the second line of \eq{xsectQ}. Explicitly, 
\bea\label{pr1}
&&\langle A_i| \Gamma^{q\bar q A}(s,\b B, \{\b b_a\};\b x,\b y) |A_i\rangle^\dagger\,
\langle A_i| \Gamma^{q\bar q A}(s,\b B, \{\b b_a\};\b x,\b y) |A_i\rangle\nonumber\\
&+& 
\langle A_i| \Gamma^{q\bar q GA}(s,\b B, \{\b b_a\};\b x, \b y, \b z_1) |A_i\rangle^\dagger\,
\langle A_i| \Gamma^{q\bar q GA}(s,\b B, \{\b b_a\};\b x, \b y, \b z_2) |A_i\rangle\nonumber\\
&-&
\langle A_i| \Gamma^{q\bar q A}(s,\b B, \{\b b_a\};\b x,\b y) |A_i\rangle^\dagger\,
\langle A_i| \Gamma^{q\bar q GA}(s,\b B, \{\b b_a\};\b x, \b y, \b z_2) |A_i\rangle\nonumber\\
&-&
\langle A_i| \Gamma^{q\bar qG A}(s,\b B, \{\b b_a\};\b x, \b y, \b z_1) |A_i\rangle^\dagger\,
\langle A_i| \Gamma^{q\bar q A}(s,\b B, \{\b b_a\};\b x,\b y) |A_i\rangle\,.
\eea

 \subsection{Incoherent diffractive gluon production} 

To calculate the propagators in the case of incoherent diffraction we need to consider each one of the four cases shown in \eq{pr1} and follow the by now familiar steps \eq{aa1}--\eq{id2}. Consider, for example, gluon emission before the interaction in the amplitude and in the c.c.\ one, i.e.\ $\tau_\sigma<\tau'_\sigma$ for $\sigma=1,2$. The detailed calculation is presented in Appendix. The result is  (cp.\ \eq{id2})
\bea
&&\sum_{f\neq i} \langle A_f| \Gamma^{q\bar q GA}(s,\b B, \{\b b_a\};\b x, \b y, \b z_1) |A_i\rangle^\dagger
\langle A_f| \Gamma^{q\bar q GA}_2(s,\b B, \{\b b_a\};\b x, \b y, \b z_2) |A_i\rangle
=\nonumber\\
&&
=
 e^{-\frac{1}{2}\left[\sigma_\mathrm{tot}^{q\bar q GN}(s;\b x, \b y, \b z_1)
+\sigma_\mathrm{tot}^{q\bar q GN}(s;\b x, \b y, \b z_2)-\frac{1}{4\pi R_p^2}
\sigma_\mathrm{tot}^{q\bar q GN}(s;\b x, \b y, \b z_1)\,\sigma_\mathrm{tot}^{q\bar q GN}(s;\b x, \b y, \b z_2)
\right]
\,\rho\, T_A(\b b)}
\nonumber\\
 && \times 
\left\{1-
e^{-\frac{1}{2}\frac{1}{4\pi R_p^2}
\sigma_\mathrm{tot}^{q\bar q GN}(s;\b x, \b y, \b z_1)\,\sigma_\mathrm{tot}^{q\bar q GN}(s;\b x, \b y, \b z_2)\,\rho\, T_A(\b b)}\right\} \label{U}\,.
\eea
This result holds in the quasi-classical approximation $\as^2A^{1/3}\sim 1$. Therefore,  contribution of elastic processes to \eq{U} are of the order
 $$
 \left(\sigma_\mathrm{tot}^{q\bar q GN}\right)^2\rho\, T_A(\b b)\sim \as^4A^{1/3}\sim \as^2\,.
$$
It is suppressed compared to the contribution of  inelastic processes which are of order $\as^2A^{1/3}\sim 1$. Note, that the contribution to the incoherent diffraction cross section   given by \eq{U} vanishes with vanishing elastic $q\bar qN$ cross section. Thus, we expand the expression in the curly brackets of \eq{U}, keeping the leading elastic term. We derive
\bea
&&\sum_{f\neq i} \langle A_f| \Gamma^{q\bar q GA}(s,\b B, \{\b b_a\};\b x, \b y, \b z_1) |A_i\rangle^\dagger
\langle A_f| \Gamma^{q\bar q GA}(s,\b B, \{\b b_a\};\b x, \b y, \b z_2) |A_i\rangle
\nonumber\\
&&\approx 
\frac{\rho\, T_A(\b b)}{8\pi R_p^2}\,
\sigma_\mathrm{tot}^{q\bar q GN}(s;\b x, \b y, \b z_1)\,\sigma_\mathrm{tot}^{q\bar q GN}(s;\b x, \b y, \b z_2)\, 
e^{-\frac{1}{2}\left[\sigma_\mathrm{tot}^{q\bar q GN}(s;\b x, \b y, \b z_1)
+\sigma_\mathrm{tot}^{q\bar q GN}(s;\b x, \b y, \b z_2)
\right]\,\rho\, T_A(\b b)}\,. \label{pr10}
\eea
This approximation is valid  at not too high energies satisfying 
\beq\label{cond1}
 \frac{1}{8\pi R_p^2}\left(\sigma_\mathrm{tot}^{q\bar qGN}\right)^2\,\rho\, T_A(\b b)\ll 1\,.
\eeq
Let us for simplicity consider a cylindrical nucleus for  which $\rho\, T_A(\b b)=2A/(\pi R_A^2)$. Since  $R_A=A^{1/3}  R_p$ and taking into account that $\frac{1}{2}\sigma_\mathrm{tot}^{q\bar qGN}\rho\, T_A(\b b)= \frac{1}{4}\,Q_s^2\, r^2$ for $N_c\gg 1$ (see \eq{sig.tot.dm} and \eq{cd2.dm})  we can re-write \eq{cond1} as
\beq\label{cond2}
\frac{1}{4}\, Q_s^2\, r^2\ll 2\, A^{1/6}\,.
\eeq
The saturation effects in the nucleus become important when $\frac{1}{4}Q_s^2r^2\gtrsim 1$. At rapidities $ Y\sim 1/\as$ the low $x$ evolution effects give rise to the energy dependence of the saturation scale.
Let $Y_1$ be the rapidity at which $\frac{1}{4}Q_s^2r^2= 1$ and $Y_2$ be the rapidity at which $\frac{1}{4}Q_s^2r^2= 2A^{1/6}$. Since   $Q_s^2\propto e^{\lambda Y}$ \cite{Levin:1999mw} we obtain $Y_2-Y_1= \lambda^{-1}\ln (2A^{1/6})$. 
Using the phenomenological value $\lambda=0.25$ for the Gold nucleus $A=197$ we find that there are $Y_2-Y_1\approx 6$ units of rapidity between $Y_1$ where the saturation effects become important and $Y_2$ where the expansion \eq{pr10} breaks down.  This provides quite wide kinematic window in which approximation \eq{cond1} holds.

Other three cases corresponding to different relations between $\tau_\sigma$ and $\tau'_\sigma$  can be worked out in a similar way.  The result for the cross section of incoherent diffractive gluon production takes form 
\bea\label{xID}
\frac{d\sigma_\mathrm{ID}(k,y)}{d^2k\, dy}&=&\frac{1}{(2\pi)^2}\,\frac{1}{8\pi R_p^2}
\int d^2b\, d^2z_1\,
d^2z_2\, \Phi^{q\bar q}(\b x, \b y, \b z_1, \b z_2)\, e^{-i\b k\cdot(\b z_1-\b z_2)}\,
\rho\, T_A(\b b)
\nonumber\\
 &  \times&
 \left( \sigma_\mathrm{tot}^{q\bar qGN}(s;\b x, \b y, \b z_1)
\,
 e^{-\frac{1}{2}\sigma_\mathrm{tot}^{q\bar qGN}(s;\b x, \b y, \b z_1)\,\rho\, T_A(\b b)}
-\sigma_\mathrm{tot}^{q\bar qN}(s;\b x, \b y)\,
e^{-\frac{1}{2}\sigma_\mathrm{tot}^{q\bar qN}(s;\b x, \b y)\,\rho\, T_A(\b b)}
 \right)
 \nonumber\\
 & \times&
 \left( \sigma_\mathrm{tot}^{q\bar qGN}(s;\b x, \b y, \b z_2)\,
 e^{-\frac{1}{2}\sigma_\mathrm{tot}^{q\bar qGN}(s;\b x, \b y, \b z_2)\,\rho\, T_A(\b b)}
- \sigma_\mathrm{tot}^{q\bar qN}(s;\b x, \b y)\,
e^{-\frac{1}{2}\sigma_\mathrm{tot}^{q\bar qN}(s;\b x, \b y)\,\rho\, T_A(\b b)}\right)\,.
\nonumber \\ 
\eea

\section{Low-$x$ evolution}\label{sec:lowx}

\subsection{Incoherent diffraction}

The results derived in the previous section can be generalized beyond the quasi-classical level to include the low-$x$ evolution. This procedure follows the general strategy developed in \cite{Kovchegov:2001sc} and has been used in \cite{Li:2008bm,Li:2008jz} to derive the expression for the coherent diffractive gluon production in onium--nucleus collisions.  In this we section we derive an analogous expressions for the incoherent diffractive gluon production. For applying this general strategy it is important, that  formula \eq{pr10} can be factorized in a product of two expressions one depending on the gluon coordinate $\b z_1$ while another one on $\b z_2$. 

The low-$x$ gluon evolution in the nucleus at large $N_c$ is taken into account by the following two substitutions in \eq{xID}: (i) the exponents are replaced by the forward elastic $q\bar q A$  scattering amplitude $N_A(\b r,\b b, Y)$   \cite{Kovchegov:2001sc}  
\beq\label{repl}
e^{-\frac{1}{2}\sigma_\mathrm{tot}^{q\bar qN}(s;\b r)\,\rho\, T_A(\b b)}\to 
1-N_A(\b r,\b b, Y)\,.
\eeq
 $N_A(\b r,\b b, Y)$ evolves towards higher rapidities $Y$, i.e.\ lower $x$, according to the nonlinear BK evolution equation \cite{Balitsky:1995ub,Kovchegov:1999yj} from its initial condition, given by \eq{sig.tot.dm} at some rapidity $Y=Y_0$;  (ii) the factors linear in $\sigma_\mathrm{tot}^{q\bar qN}$ have emerged in expansion \eq{pr10} where the terms of higher order (i.e.\ multiple scattering) in the elastic amplitude were neglected. Therefore, they are replaced by the forward elastic $q\bar qN$  scattering amplitude $N_p(\b r, \b b, Y)$ that evolves according to the linear BFKL equation \cite{Kuraev:1977fs,Balitsky:1978ic}. It is defined similarly to  \eq{NA}  (see \eq{bpp})
\beq\label{Np}
N_p(\b r, \b b, Y)=\mathrm{Re}\, \Gamma^{q \bar qp}(s,\b b;\b r)=\frac{1}{2}\sigma_\mathrm{tot}^{q\bar q p}(s;\b r)\,S_p(\b b)\,.
\eeq
Recall, that  in the heavy nucleus environment the impact parameter dependence of $q\bar qp$ cross section can be neglected, see \eq{overlap}. Thus, using \eq{sp} we have
\beq\label{W}
\sigma_\mathrm{tot}^{q\bar q N}(s;\b r)\to 2\pi R_p^2\, N_p(\b r,  0, Y)\,
\eeq
in agreement with the optical theorem.
The factors linear in the $q\bar qGN$ cross section are replaced similarly using \eq{expr}.  Substituting \eq{repl} and \eq{W} into \eq{xID} using \eq{expr} (in the multicolor  $N_c\gg 1$ limit) yields
\bea\label{xsectID}
&&\frac{d\sigma_\mathrm{ID}(k,y)}{d^2k\, dy}=\frac{1}{(2\pi)^2}\,\frac{\pi R_p^2}{2}
\int d^2b\, d^2z_1\,
d^2z_2\, \Phi^{q\bar q}(\b x, \b y, \b z_1, \b z_2)\,
\rho\, T_A(\b b)\,
e^{-i\b k\cdot(\b z_1-\b z_2)}\,
\nonumber\\
 &&\times\,
 \bigg\{ 
 \left[ 1-N_A(\b z_1-\b x,\b b, y) \right]\, \left[ 1-N_A(\b z_1-\b y,\b b, y) \right]
 \, \left[ N_p(\b z_1-\b x,0, y) +  N_p(\b z_1-\b y,0, y)  \right]
 \nonumber\\
 &&
 -   \left[ 1-N_A(\b x-\b y,\b b, y) \right] N_p(\b x-\b y,0, y) 
 \bigg\}
 \nonumber\\
 &&
\times\,
 \bigg\{ 
 \left[ 1-N_A(\b z_2-\b x,\b b, y) \right]\, \left[ 1-N_A(\b z_2-\b y,\b b, y) \right]
 \, \left[ N_p(\b z_2-\b x,0, y) +  N_p(\b z_2-\b y,0, y)  \right]
 \nonumber\\
 &&
 -   \left[ 1-N_A(\b x-\b y,\b b, y) \right] N_p(\b x-\b y,0, y) 
 \bigg\}\,.
\eea

We can write \eq{xsectID} in a more compact form introducing the following two-dimensional vector
\bea\label{iid}
&&\b I_\mathrm{ID}(\b x-\b y,\b b, y;\b k)= \int d^2 z \left(\frac{\b z-\b x}{|\b z-\b x|^2}-
 \frac{\b z-\b y}{|\b z-\b y|^2}\right)\,e^{-i\b k \cdot \b z}\,\nonumber\\
 && 
 \times\,
 \bigg\{ 
 \left[ 1-N_A(\b z-\b x,\b b, y) \right]\, \left[ 1-N_A(\b z-\b y,\b b, y) \right]
 \, \left[ N_p(\b z-\b x,0, y) +  N_p(\b z-\b y,0, y)  \right]
 \nonumber\\
 &&
 -   \left[ 1-N_A(\b x-\b y,\b b, y) \right] N_p(\b x-\b y,0, y) 
 \bigg\}\,.
\eea
Then,
\beq\label{comp1}
\frac{d\sigma_\mathrm{ID}(k,y)}{d^2k\, dy}=\frac{\as C_F}{\pi^2}\, \frac{1}{(2\pi)^2}\,\frac{\pi R_p^2}{2}\,
\int d^2b\, \rho\, T_A(\b b)\, |\b I_\mathrm{ID}(\b x-\b y,\b b, y;\b k)|^2\,.
\eeq

So far we have discussed the gluon evolution in the rapidity interval $y$ between the emitted gluon and the target nucleus. The low-$x$ evolution also occurs in the rapidity interval $Y-y$ between the incident quarkonium $q\bar q$ and the emitted gluon (we now denote by $Y$  the rapidity between the quarkonium and the nucleus and by $y$ the inclusive gluon rapidity).  The low-$x$ evolution in the incident quarkonium is taken into account by convoluting \eq{comp1} with the dipole density $n_1(\b r,\b r', \b B-\b b, Y-y)$, where $\b r=\b x-\b y$:
\beq\label{all1}
\frac{d\sigma_\mathrm{ID}(k,y)}{d^2k\, dy}=\frac{\as C_F}{\pi^2}\, \frac{\pi R_p^2}{2(2\pi)^2}
\int d^2b \int d^2B \int d^2r'\, n_1(\b r, \b r', \b B-\b b, Y-y)\, \rho\, T_A(\b b)\, |\b I_\mathrm{ID}(\b r,\b b, y;\b k)|^2\,.
\eeq
The dipole density $n_1(\b r,\b r,\b B-\b b,Y-y)$ has the meaning of the number of dipoles of size $\b r'$ at rapidity $Y-y$ and impact parameter $\b b$ generated by evolution from the original dipole $\b r$ having rapidity $Y$ and impact parameter $\b B$ \cite{dip}. It obeys the BFKL equation \cite{Kuraev:1977fs,Balitsky:1978ic}. As the interval  $Y-y$ 
increases the quarkonium wave function involves increasing number of dipoles. 
Although the dipole size distribution diffuses according to the BFKL equation, the typical dipole size is much smaller than the nuclear radius $R_A$, i.e.\ $|\b B-\b b|\ll b$. Integrating over $\b B-\b b$ in this approximation we derive 
\beq\label{all2}
\frac{d\sigma_\mathrm{ID}(k,y)}{d^2k\, dy}=\frac{\as C_F}{\pi^2}\, \frac{\pi R_p^2}{2(2\pi)^2}
\int d^2b  \int d^2r'\, n_p(\b r, \b r', Y-y)\, \rho\, T_A(\b b)\, |\b I_\mathrm{ID}(\b r,\b b, y;\b k)|^2\,,
\eeq
where we defined \cite{Li:2008bm,Li:2008jz}
\beq\label{np20}
n_p(\b r, \b r', Y)=\int n_1(\b r, \b r', \b b', Y)\, d^2b'\,.
\eeq
Eqs.~\eq{all2} and \eq{iid} constitute the main result of this paper.

\subsection{Coherent diffraction}

The cross section for the coherent diffractive gluon production including the low-$x$ evolution was derived in \cite{Li:2008bm,Li:2008jz}. It can be written similarly to  \eq{all2} \cite{Li:2008bm,Li:2008jz}:
\beq\label{cd10}
\frac{d\sigma_\mathrm{CD}(k,y)}{d^2k\, dy}=\frac{\as C_F}{\pi^2}\, \frac{1}{(2\pi)^2}
\int d^2b  \int d^2r'\, n_p(\b r, \b r', Y-y)\, |\b I_\mathrm{CD}(\b r,\b b, y;\b k)|^2\,,
\eeq
where
\bea\label{icd}
\b I_\mathrm{CD}(\b x-\b y,\b b, y;\b k)&=& \int d^2 z \left(\frac{\b z-\b x}{|\b z-\b x|^2}-
 \frac{\b z-\b y}{|\b z-\b y|^2}\right)\,e^{-i\b k \cdot \b z}\,\nonumber\\
 && 
 \times\,
 \bigg\{ 
 -N_A(\b z-\b x,\b b, y) -N_A(\b z-\b y,\b b, y)  +N_A(\b x-\b y,\b b, y)  \nonumber\\
 &&
 +N_A(\b z-\b x,\b b, y) \,N_A(\b z-\b y,\b b, y)
  \bigg\}\,.
\eea
In \cite{Li:2008bm,Li:2008jz} we presented a detailed analytical and numerical analysis of the the coherent diffractive gluon production. 

\subsection{ Logarithmic approximation}

To obtain a working model for numerical calculations, it is useful  to estimate $\b I_\mathrm{ID}(\b r,\b b, y;\b k)$ in the logarithmic approximation. First, let us change the integration variable $\b w= \b z-\b y$ and define an auxiliary function \cite{Li:2008bm,Li:2008jz}
\bea\label{QID}
Q_\mathrm{ID}(\b r', y;\b k)&=&\int \frac{d^2w}{w^2}\, e^{i\b k\cdot \b w}\,
 \bigg\{ \left[ 1-N_A(\b r',\b b, y) \right] N_p(\b r',0, y)  \nonumber\\
 &-&  
 \left[ 1-N_A(\b w-\b r',\b b, y) \right]\, \left[ 1-N_A(\b w,\b b, y) \right]
 \, \left[ N_p(\b w-\b r',0, y) +  N_p(\b w,0, y)  \right]
 \bigg\}\,.
\eea
With this definition  \eq{iid} becomes 
\beq\label{bb1}
\b I_\mathrm{ID}(\b r,\b b, y;\b k)=-e^{-i\b k\cdot \b r}\,i\nabla_{\b k}Q_\mathrm{ID}(\b r', y;\b k)+e^{i\b k\cdot \b r}\,i\nabla_{\b k}Q_\mathrm{ID}^*(\b r', y;\b k)\,.
\eeq
In the logarithmic approximation $Q_\mathrm{ID}(\b r', y;\b k)$ is a real function\cite{Li:2008bm,Li:2008jz}. Therefore,  we can write 
\beq\label{abs}
|\b I_\mathrm{ID}(\b r,\b b, y;\b k)|^2= 4\sin^2\big(\frac{1}{2}\,\b k\cdot \b r'\big)\, (\nabla_{\b k}Q_\mathrm{ID}(\b r', y;\b k))^2\,.
\eeq
Notice that the integral over $\b z$  in \eq{iid} is dominated by the region $w<1/k$ since otherwise the integrand is a rapidly oscillation function. Restricting integration to this region yields results valid as long as the scales $1/r'$, $Q_s$ and $k$ are strongly ordered \cite{Li:2008bm,Li:2008jz}. Operating with $\nabla_{\b k}$ on $Q_\mathrm{ID}(\b r', y;\b k)$ yields
\bea\label{log1}
&&\nabla_{\b k}Q_\mathrm{ID}(\b r', y;\b k)= -\frac{\b k}{k^3}\frac{\partial }{\partial k^{-1}}Q_\mathrm{ID}(\b r', y;\b k)=
 -2\pi\frac{\b k}{k^2}\bigg\{ \left[ 1-N_A(\b r',\b b, y) \right] N_p(\b r',0, y) 
\nonumber\\
&& -  
 \left[ 1-N_A(\b k\,k^{-2}-\b r',\b b, y) \right]\, \left[ 1-N_A(\b k\,k^{-2},\b b, y) \right]
 \, \left[ N_p( \b k\,k^{-2}-\b r',0, y) +  N_p(\b k\,k^{-2},0, y)  \right]
 \bigg\}\,.
\eea

Analogously, in the coherent diffraction case we define
\bea\label{QCD}
Q_\mathrm{CD}(\b r', y;\b k)&=&\int \frac{d^2w}{w^2}\, e^{i\b k\cdot \b w}\,
 \bigg\{
N_A(\b w-\b r',\b b, y) +N_A(\b w,\b b, y) -N_A(\b r',\b b, y)
 \nonumber\\
 &&   -N_A(\b w-\b r',\b b, y)\,N_A(\b w,\b b, y) 
    \bigg\}\,.
\eea
Then, in the logarithmic approximation, 
\bea\label{log2}
\nabla_{\b k}Q_\mathrm{CD}(\b r', y;\b k)&=&
 -2\pi\frac{\b k}{k^2}\bigg\{ -N_A(\b r',\b b, y) +N_A(\b k\,k^{-2}-\b r',\b b, y)+
N_A(\b k\,k^{-2},\b b, y)\nonumber\\
&&
-N_A(\b k\,k^{-2}-\b r',\b b, y)\,N_A(\b k\,k^{-2},\b b, y)
\bigg\}\,.
\eea
Similarly \eq{abs}  becomes
\beq\label{abs1}
|\b I_\mathrm{CD}(\b r,\b b, y;\b k)|^2= 4\sin^2\big(\frac{1}{2}\,\b k\cdot \b r'\big)\, (\nabla_{\b k}Q_\mathrm{CD}(\b r', y;\b k))^2\,.
\eeq

\section{Numerical calculations} \label{sec:nmf}

In  our previous paper \cite{Li:2008se} we justified using the quarkonium--nucleus $q\bar qA$ scattering  as a model of $pA$ collisions. It also serves as the QCD ingredient for the deep inelastic $eA$ scattering.  Therefore, we extend the use of this model in this section for the incoherent diffraction. 

We fix the quarkonium size at $r=0.2$~fm. Dependence of all cross sections on $r$ has been discussed at length in \cite{Li:2008bm,Li:2008jz,Li:2008se} where  the interested reader is readily referred. Since our main goal in this section is to illustrate the key features of the various diffraction channels, rather than giving a detailed quantitative analyses, we approximate the nuclear profile by the step function (cylindrical nucleus). 

It is clear from the discussion in the previous Section that in order to calculate the diffractive cross sections we need only to specify the dipole--nucleus scattering amplitude $N_A(\b r,\b b, y)$ and the dipole density $n_p(\b r, \b r', y)$. The  dipole--proton scattering amplitude $N_p(\b r,\b b, y)$ is obtained from $N_A(\b r,\b b, y)$ in the limit $A\to 1$. 

For the forward elastic dipole--nucleus scattering amplitude $N_A(\b r, \b b, y)$ we use the KKT model \cite{Kharzeev:2004yx}, which is based on analytical analysis in \cite{Kharzeev:2003wz}. It successfully describes the inclusive hadron production in the RHIC kinematic region (at the central and forward rapidities). 

For the dipole density we use the leading BFKL solution in the diffusion approximation: 
\beq\label{ndiff}
n_p( r, r', Y-y)=\frac{1}{2\pi^2}\frac{1}{rr'}\sqrt{\frac{\pi}{14\zeta(3)\bas\,d\, (Y-y)}}\, e^{(\alpha_P-1)(Y-y)}\, e^{-\frac{\ln^2\frac{r}{r'}}{14\zeta(3)\bas\, d\, (Y-y)}}\,.
\eeq
Parameter $d$ is equal to unity in the LO BFKL. To obtain the hadron diffractive cross section we convoluted the gluon cross sections  \eq{all2} and \eq{cd10} with the LO pion fragmentation function given in \cite{Kniehl:2000hk}.

\begin{figure}[ht]
\begin{tabular}{cc}
      \includegraphics[width=8cm]{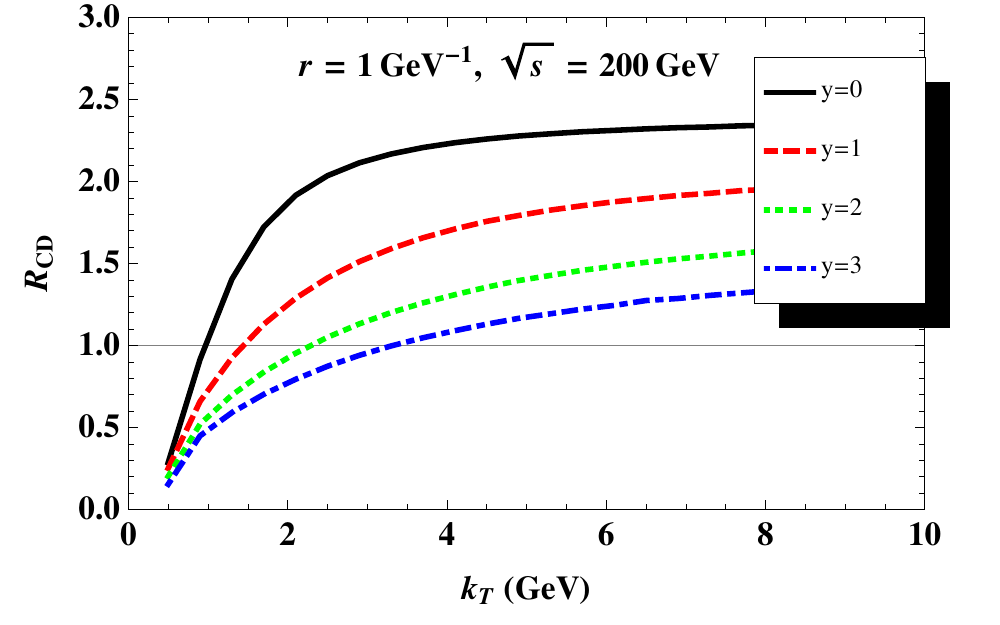} &
      \includegraphics[width=8cm]{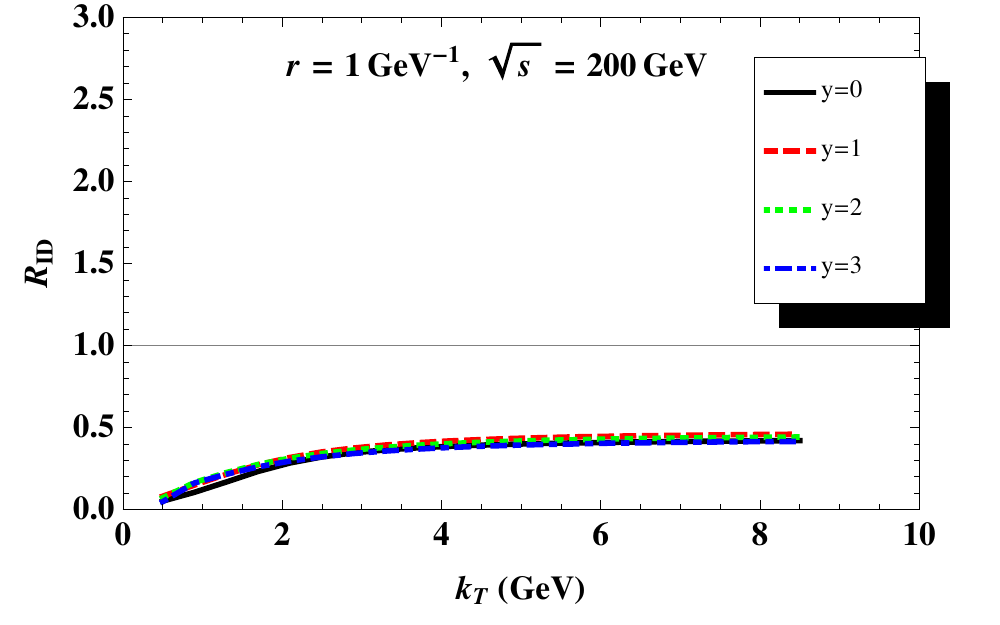}\\
      (a) & (b)
\end{tabular}
\caption{(Color online). Nuclear modification factor for (a) coherent diffraction, (b) incoherent diffraction for RHIC at different rapidities. }
\label{cd+id}
\end{figure}

A convenient way to study the nuclear dependence of  particle production is to consider  the nuclear modification factor defined as follows
\beq\label{nmf}
R^{pA}_\mathrm{diff}(k_T,y)=\frac{\frac{d\sigma^{pA}_\mathrm{diff}(k_T,y)}{d^2k_Tdy}}{A\,\frac{d\sigma^{pp}_\mathrm{diff}(k_T,y)}{d^2k_Tdy} }\,.
\eeq
We calculate the diffractive gluon production in $pp$ collisions, which is required as a baseline for the calculation of the nuclear modification factor \eq{nmf},  by setting $A=1$ in the formula for the corresponding cross section in $pA$ collisions. The nuclear modification factor is defined in such a way that a completely incoherent \emph{scattering} would yield  $R^{pA}_\mathrm{diff}(k_T,y)=1$. The results of the calculations are exhibited in \fig{cd+id}
We see that the nuclear modification factor is rapidly decreasing with rapidity  for the coherent diffraction case and is almost independent of rapidity for the incoherent case. This trend continues even at the forward LHC energies \cite{Li:2008se}. In contrast to inclusive hadron production that saturates already in the forward rapidities at RHIC \cite{Kharzeev:2004yx,Tuchin:2007pf},  we expect that the coherent diffraction cross section   saturates at much higher energies and rapidities, perhaps at the forward rapidities  at LHC (see \cite{Li:2008se} for an extensive discussion). As \fig{cd+id} implies, the incoherent diffractive gluon production is saturated already in the central rapidity region at RHIC. 

\begin{figure}[ht]
\begin{tabular}{cc}
      \includegraphics[width=8cm]{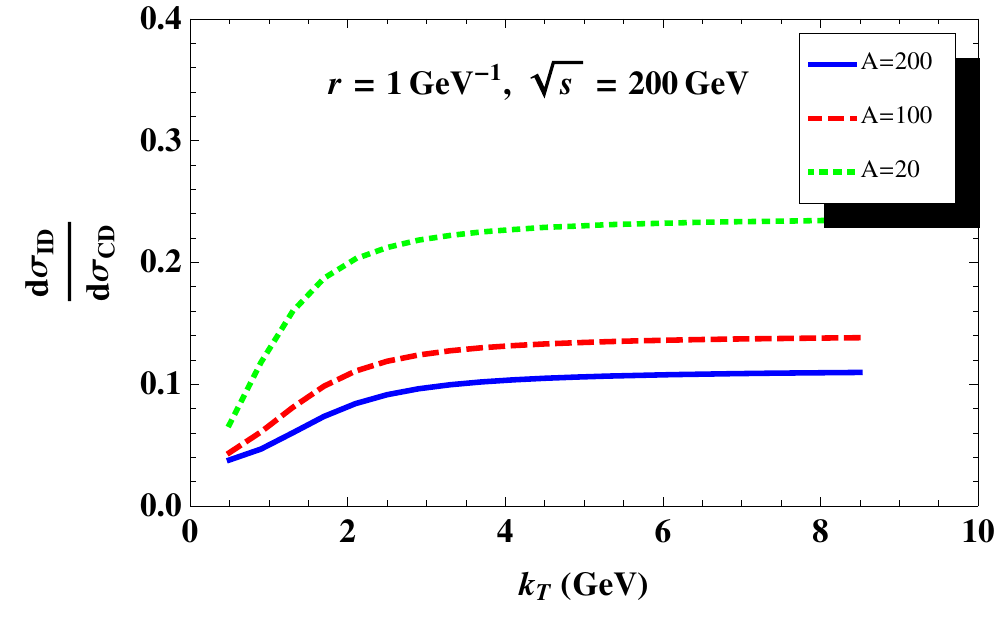} &
      \includegraphics[width=8cm]{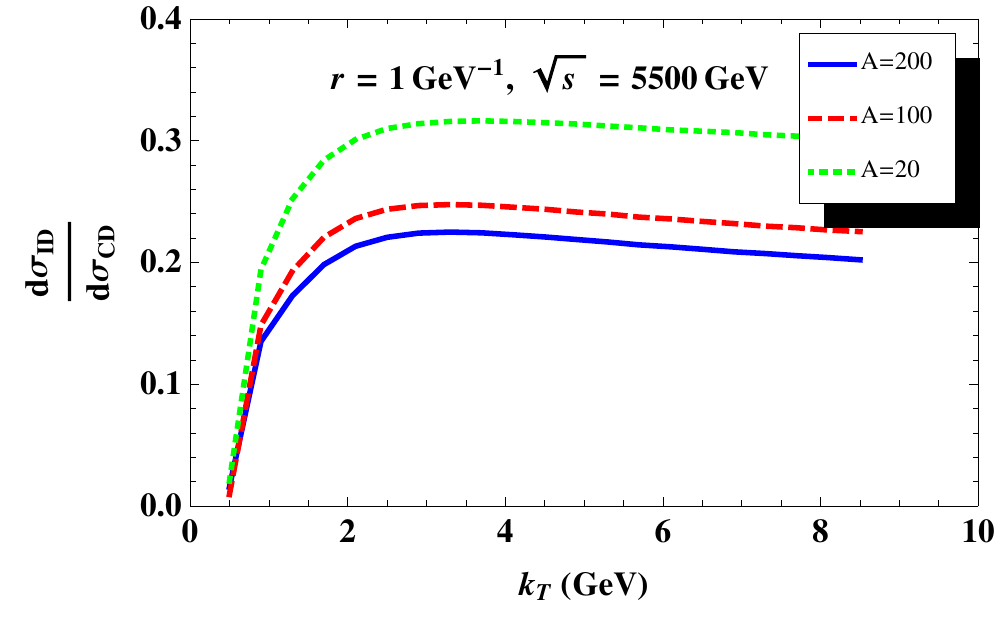} \\
      (a) & (b)
\end{tabular}
\caption{(Color online). Ratio of the cross sections for coherent and incoherent diffractive gluon production in $pA$ collisions  at midrapidity: (a) RHIC and (b) LHC. Solid blue line: $A=200$, dashed red line: $A=100$ and dotted green line: $A=20$. }
\label{fig:rat} 
\end{figure}

We observe in \fig{cd+id} a parametric enhancement by a factor  $\sim A^{1/3}$  of the coherent diffractive cross section with respect to the incoherent one. This is a benchmark of the classical gluon field of the nucleus in which quantum fluctuations are suppressed by  $A^{1/3}$.
 This feature is also seen in \fig{fig:rat} where we show the ratio of the cross sections for coherent and incoherent diffractive gluon production for $pA$ collisions at different energies. To illustrate the atomic number dependence of this ratio we 
show three cases: $A=200$, $A=100$ and $A=20$. 

In \fig{fig:rat} we can see that the fraction of the incoherent diffractive events increases from RHIC to LHC. This happens because the gluon saturation effects in proton are still not strong enough to unitarize the cross section. Remember, this allowed us to expand  the scattering amplitude as shown in \eq{pr10}. Our estimate \eq{cond2} shows that this approximation is valid for about six units of  rapidity starting from the rapidity $y_1\simeq 1$ at RHIC. Therefore, our conclusion that the fraction of incoherent diffractive events increases holds up to the rapidity $y_2\simeq 3$ at the LHC. As soon as the gluon distribution of the incident  proton saturates, the incoherent diffraction cross section vanishes, as can be seen in \eq{id2}. This effect however is not included in our present model. 

\begin{figure}[ht]
\begin{tabular}{cc}
      \includegraphics[width=8cm]{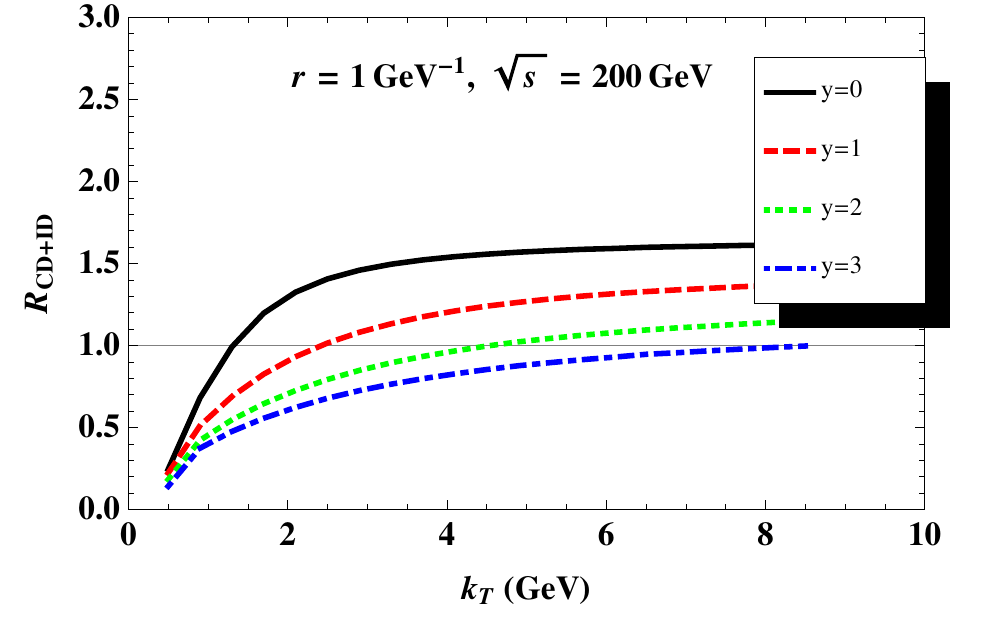} &
      \includegraphics[width=8cm]{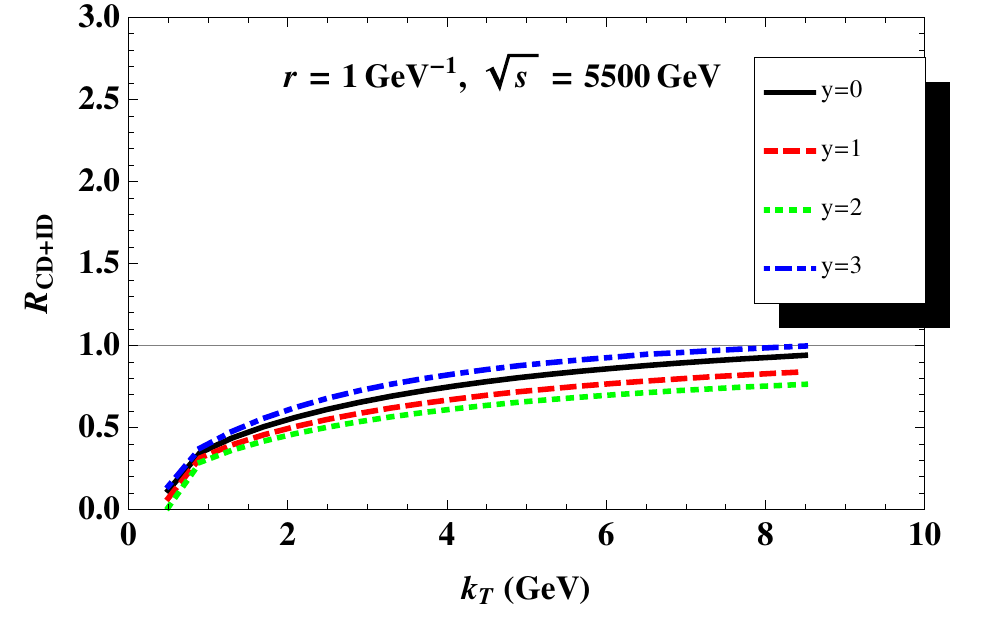}\\
      (a) & (b)
\end{tabular}
\caption{(Color online). Nuclear modification factor for the sum of coherent and incoherent diffractive cross sections  for (a) RHIC, (b) LHC at different rapidities. }
\label{fig:tot}
\end{figure}

Finally, it may be challenging to experimentally distinguish  the
coherent and incoherent diffractive cross sections. In part it is related to the difficulty of performing measurements at very forward angles. One usually measures an event with large rapidity gap between the produced hadronic system and the target remnants in the forward direction (that may or may not be an intact nucleus). In such a likely case we consider the nuclear modification factor for a sum of coherent and incoherent diffractive channels. The result is 
displayed in \fig{fig:tot}. We can see that the predictions for the total diffractive cross section are similar to those of the inclusive hadron production \cite{Tuchin:2007pf}. Therefore, it is very important to experimentally separate contributions of coherent and incoherent diffraction. The former is especially useful for studying the low $x$ QCD due to its strong and non-trivial energy, rapidity and atomic number dependence.  

Let us also mention that electromagnetic interactions also contribute to diffraction processes in $pA$ collisions. If they become large, as argued in  \cite{Guzey:2008gw}, it may present an additional experimental challenge. We intend to analyze contribution of the e.m.\ interactions in a separate publication.

To conclude this section, we would like to emphasize that  Figs.~\ref{cd+id},\ref{fig:rat},\ref{fig:tot} are an \emph{illustration} of  general features expected in various diffractive channels. As we have mentioned more than once in this paper, application of  realistic experimental cuts, which are unique for every experiment, will significantly change the absolute values of the cross sections and relative importance of the coherent and the incoherent channels. A dedicated study is required in each case, although the key features displayed in Figs.~\ref{cd+id},\ref{fig:rat},\ref{fig:tot}  will perhaps remain unchanged.

\section{Summary}\label{sec:summary}

To summarize, we calculated the cross section for the incoherent diffractive gluon production in $q\bar qA$ collisions, which is a prototype of $pA$ collisions at RHIC and LHC and $\gamma^*A$ ones at EIC. We  took into account the gluon saturation, i.e.\ color glass condensate, effects. We then compared prediction of diffractive hadron production in coherent (calculated previously in \cite{Li:2008bm,Li:2008jz,Li:2008se}) and incoherent diffraction. 

Coherent diffractive gluon production has a very characteristic energy, rapidity  and atomic number dependence which makes it a powerful tool for studying the leading mean-field contribution to the color glass condensate. 
Incoherent diffraction arises from contributions to the gluon field correlations beyond the mean-field approximation. Its experimental study can reveal the 
dynamics of quantum fluctuations in the CGC. Unlike the nuclear modification factor for coherent diffractive gluon production the nuclear modification factor for incoherent diffraction is not expected to exhibit a significant rapidity and energy dependence. Ratio of the coherent and incoherent inclusive diffractive cross sections is predicted to increase from RHIC to LHC \emph{if} the gluon saturation effects in proton are small. Otherwise, the ratio will decrease approaching the unitarity limit (i.e.\ zero).  

Finally,  one may consider measuring a diffractive  event  with  large rapidity gap but without distinguishing contributions of coherent and incoherent components. Although in this case many interesting features of coherent and incoherent channels get averaged out, the corresponding cross section as well as the nuclear modification factor are expected to display a non-trivial behavior as a function of energy, atomic number and transverse momentum. This behavior is sensitive to the underlying parton dynamics and thus can serve as a discriminator  of different models. We are certain that 
studying  diffraction at RHIC, LHC and EIC can become an important tool in accessing the detailed structure of QCD at low-$x$.

\acknowledgments
I would like to thank Dima Kharzeev, Yura Kovchegov  and J.-W.~Qiu for informative discussions.  I am grateful to Wlodek Guryn for encouragement and for explaining to me numerous experimental challenges in measurements of diffractive processes at RHIC. 
This work  was supported in part by the U.S. Department of Energy under Grant No.\ DE-FG02-87ER40371. I would like to
thank RIKEN, BNL, and the U.S. Department of Energy (Contract No.\ DE-AC02-98CH10886) for providing facilities essential
for the completion of this work.

\appendix
\section{}

In this Appendix we derive Eq.~\eq{U}. For notation simplicity we replace the coordinates $\b x, \b y, \b z_\sigma$ in the scattering amplitudes by  subscript $\sigma$, with $\sigma=1,2$. We have
\bea\label{pr2}
&&\sum_{f\neq i} \langle A_f| \Gamma^{q\bar q GA}_1(s,\b B, \{\b b_a\}) |A_i\rangle^\dagger
\langle A_f| \Gamma^{q\bar q GA}_2(s,\b B, \{\b b_a\}) |A_i\rangle
=
\nonumber\\
& & 
  \langle A_i|\Gamma^{q\bar qGA}_1(s,\b B, \{\b b_a\})\,\Gamma^{q\bar qGA}_2(s,\b B, \{\b b_a\})|A_i\rangle\nonumber\\
&&
- \langle A_i|\Gamma^{q\bar qGA}_1(s,\b B, \{\b b_a\})|A_i\rangle
 \langle A_i|\Gamma^{q\bar qGA}_2(s,\b B, \{\b b_a\})|A_i\rangle
\,,
 \eea
Similarly to \eq{aa2} we write
\bea\label{pr3}
&&\Gamma^{q\bar qGA}_1(s,\b B, \{\b b_a\})\,\Gamma^{q\bar qGA}_2(s,\b B, \{\b b_a\})\,=\,
1-\prod_{a=1}^A\left[  1- \Gamma^{q\bar qGN}_1(s,\b B-\b b_a)\right] 
\nonumber\\
&& 
-\prod_{a=1}^A\left[  1- \Gamma^{q\bar qGN}_2(s,\b B-\b b_a)\right]
+
\prod_{a=1}^A \left[  1- \Gamma^{q\bar qGN}_1(s,\b B-\b b_a)\right]
\left[  1- \Gamma^{q\bar qGN}_2(s,\b B-\b b_a)\right]
\eea
Averaging over the nucleus
\bea\label{pr5}
&&
\langle A_i|   \Gamma^{q\bar qGA}_1(s,\b B, \{\b b_a\})\,\Gamma^{q\bar qGA}_2(s,\b B, \{\b b_a\})    |A_i\rangle
\nonumber\\
&&=1-e^{-\int d^2b\, \Gamma^{q\bar qGN}_1(s,\b B-\b b)\rho\, T_A(b)}
-\,e^{-\int d^2b\, \Gamma^{q\bar qGN}_2(s,\b B-\b b)\rho\, T_A(b)}
\nonumber\\
&&
+
e^{-\int d^2b\, \left[\Gamma^{q\bar q GN}_1(s,\b B-\b b)+\Gamma^{q\bar q GN}_2(s,\b B-\b b)-\Gamma^{q\bar q GN}_1(s,\b B-\b b)\Gamma^{q\bar q GN}_2(s,\b B-\b b)\right] \rho\, T_A(b)}
\eea
and subtracting the coherent part
\beq\label{pr50}
\left( 1-\Gamma^{q\bar q GA}_1(s,\b B, \{\b b_a\})\right) \, \left( 1-\Gamma^{q\bar q GA}_2(s,\b B, \{\b b_a\})\right) 
\eeq
we  obtain
\bea\label{pr6}
&&\sum_{f\neq i} \langle A_f| \Gamma^{q\bar q GA}_1(s,\b B, \{\b b_a\}) |A_i\rangle^\dagger
\langle A_f| \Gamma^{q\bar q GA}_2(s,\b B, \{\b b_a\}) |A_i\rangle
=\nonumber\\
&&
e^{-\int d^2b\, \left[\Gamma^{q\bar q GN}_1(s,\b B-\b b)+\Gamma^{q\bar q GN}_2(s,\b B-\b b)-\Gamma^{q\bar q GN}_1(s,\b B-\b b)\Gamma^{q\bar q GN}_2(s,\b B-\b b)\right] \rho\, T_A(b)}\nonumber\\
&&
-e^{-\int d^2b\, \left[\Gamma^{q\bar qGN}_1(s,\b B-\b b)+\Gamma^{q\bar qGN}_2(s,\b B-\b b)\right]\rho\, T_A(b)}\\
&&=
e^{-\int d^2b\, \left[\Gamma^{q\bar q GN}_1(s,\b B-\b b)+\Gamma^{q\bar q GN}_2(s,\b B-\b b)-\Gamma^{q\bar q GN}_1(s,\b B-\b b)\Gamma^{q\bar q GN}_2(s,\b B-\b b)\right] \rho\, T_A(b)}\nonumber\\
&&\times
\left\{ 
1-e^{-\int d^2b\, \Gamma^{q\bar qGN}_1(s,\b B-\b b)\Gamma^{q\bar qGN}_2(s,\b B-\b b)\rho\, T_A(b)}
\right\}\\
&&= e^{-\frac{1}{2}\left[\sigma_\mathrm{tot}^{q\bar q GN}(s,1)
+\sigma_\mathrm{tot}^{q\bar q GN}(s,2)-\frac{1}{4\pi R_p^2}
\sigma_\mathrm{tot}^{q\bar q GN}(s,1)\,\sigma_\mathrm{tot}^{q\bar q GN}(s,2)
\right]
\,\rho\, T_A(\b b)}
\nonumber\\
&&\times \left\{1-
e^{-\frac{1}{2}\frac{1}{4\pi R_p^2}
\sigma_\mathrm{tot}^{q\bar q GN}(s,1)\,\sigma_\mathrm{tot}^{q\bar q GN}(s,2)\,\rho\, T_A(\b b)}\right\} \,,
\eea
as advertised. In the last line we employed the approximation $R_p\ll R_A$ which allows to write using an analogue of \eq{bpp} 
\bea\label{appr1}
&&\int d^2b \, \Gamma^{q\bar qGN}_1(s,\b B-\b b)\Gamma^{q\bar qGN}_2(s,\b B-\b b)\,\rho\, T_A(b)\nonumber\\
&&
= \frac{1}{4}\sigma_\mathrm{tot}^{q\bar q GA}(s,1)\,\sigma_\mathrm{tot}^{q\bar q GA}(s,2)\,\int d^2b\, \frac{1}{(\pi R_p^2)^2}\, e^{-2\frac{(\b b-\b B)^2}{R_p^2}}\,\rho\, T_A(b)\nonumber\\
&&\approx
\frac{1}{4}\sigma_\mathrm{tot}^{q\bar q GA}(s,1)\,\sigma_\mathrm{tot}^{q\bar q GA}(s,2)\,\frac{1}{(\pi R_p^2)^2}\,\rho\, T_A(\b B) \int d^2b'\, 
e^{-2\frac{ b'^2}{R_p^2}}\nonumber\\
&&=
\frac{1}{4}\sigma_\mathrm{tot}^{q\bar q GA}(s,1)\,\sigma_\mathrm{tot}^{q\bar q GA}(s,2)\,\frac{1}{2\pi R_p^2}\,\rho\, T_A(\b B)\,.
\eea
where $\b b'=\b b -\b B$ and  $\b b'^2\ll \b b^2,\, \b B^2$.


\end{document}